
\catcode`@=11

\newskip\ttglue

\font\twelverm=cmr12 \font\twelvebf=cmbx12
\font\twelveit=cmti12 \font\twelvesl=cmsl12

\font\ninerm=cmr9
\font\eightrm=cmr8
\font\sixrm=cmr6
\font\eighti=cmmi8   \skewchar\eighti='177
\font\sixi=cmmi6     \skewchar\sixi='177
\font\ninesy=cmsy9   \skewchar\ninesy='60
\font\eightsy=cmsy8  \skewchar\eightsy='60
\font\sixsy=cmsy6    \skewchar\sixsy='60
\font\eightbf=cmbx8
\font\sixbf=cmbx6
\font\eighttt=cmtt8  \hyphenchar\eighttt=-1
\font\eightit=cmti8
\font\eightsl=cmsl8

\def\smalltype{\def\rm{\fam0\eightrm}
 			\textfont0=\eightrm  \scriptfont0=\sixrm  \scriptscriptfont0=\fiverm
 			\textfont1=\eighti   \scriptfont1=\sixi   \scriptscriptfont1=\fivei
 			\textfont2=\eightsy  \scriptfont2=\sixsy  \scriptscriptfont2=\fivesy
 			\textfont3=\tenex    \scriptfont3=\tenex  \scriptscriptfont3=\tenex
    \textfont\itfam=\eightit  \def\it{\fam\itfam\eightit}
	   \textfont\slfam=\eightsl  \def\sl{\fam\slfam\eightsl}
	   \textfont\ttfam=\eighttt  \def\tt{\fam\ttfam\eighttt}
    \textfont\bffam=\eightbf  \scriptfont\bffam=\sixbf
        \scriptscriptfont\bffam=\fivebf  \def\bf{\fam\bffam\eightbf}
    \tt  \ttglue=.5em plus.25em minus.15em
    \normalbaselineskip=9pt
    \setbox\strutbox=\hbox{\vrule height7pt depth2pt width0pt}
    \let\sc=\sixrm  \let\big=\eightbig  \normalbaselines\rm}
\def\eightbig#1{{\hbox{$\textfont0=\ninerm\textfont2=\ninesy
    \left#1\vbox to6.5pt{}\right.\n@space$}}}

\def\medtype{\def\rm{\fam0\tenrm}
 			\textfont0=\tenrm  \scriptfont0=\sevenrm  \scriptscriptfont0=\fiverm
 			\textfont1=\teni   \scriptfont1=\seveni   \scriptscriptfont1=\fivei
 			\textfont2=\tensy  \scriptfont2=\sevensy  \scriptscriptfont2=\fivesy
 			\textfont3=\tenex    \scriptfont3=\tenex  \scriptscriptfont3=\tenex
    \textfont\itfam=\tenit  \def\it{\fam\itfam\tenit}
	   \textfont\slfam=\tensl  \def\sl{\fam\slfam\tensl}
	   \textfont\ttfam=\tentt  \def\tt{\fam\ttfam\tentt}
    \textfont\bffam=\tenbf  \scriptfont\bffam=\sevenbf
        \scriptscriptfont\bffam=\fivebf  \def\bf{\fam\bffam\tenbf}
    \tt  \ttglue=.5em plus.25em minus.15em
    \normalbaselineskip=12pt
    \setbox\strutbox=\hbox{\vrule height8.5pt depth3.5pt width0pt}
    \let\sc=\eightrm  \let\big=\tenbig  \normalbaselines\rm}

\def\bigtype{\let\rm=\twelverm \let\bf=\twelvebf
\let\it=\twelveit \let\sl=\twelvesl \rm}

\def\footnote#1{\edef\@sf{\spacefactor\the\spacefactor}#1\@sf
    \insert\footins\bgroup\smalltype
    \interlinepenalty100 \let\par=\endgraf
    \leftskip=0pt  \rightskip=0pt
    \splittopskip=10pt plus 1pt minus 1pt \floatingpenalty=20000
  \vskip4pt\noindent\hskip20pt\llap{#1\enspace}
\bgroup\strut\aftergroup\@foot\let\next}
\skip\footins=12pt plus 2pt minus 4pt \dimen\footins=30pc

\def\bigfont{\magnification=1200 \baselineskip=20pt}

\def\a{\alpha}  \def\d{\delta}
\def\e{\epsilon}  
\def\s{\sigma} \def\t{\tau}

\def\cl#1{\centerline{#1}}
\def\clbf#1{\centerline{\bf #1}}

\def\is#1{{\narrower\smallskip\noindent#1\smallskip}}

\def\ve{\vfill\eject}

\def\frac#1#2{{#1 \over #2}}
\def\Re{I\!\!R}

\newcount\sectnumber
\def\Section#1{\global\advance\sectnumber by 1 \bigskip
           \cl{\bigtype \the\sectnumber. #1} \medskip}

\def\defn#1{\medskip\noindent {\bf Definition #1. }}

\def\lemma#1{\medskip\noindent {\bf Lemma #1.} \it}
\def\thm#1{\medskip\noindent {\bf Theorem #1.} \it}

\def\ok{\smallskip \rm}

\def\pf{\medskip\noindent Proof: \/}
\def\pfo#1{\medskip\noindent {\bf Proof of #1:\/}}
\def\endpf{\hfill {\it Q.E.D.} \smallskip}

\newcount\notenumber
\def\note#1{\global\advance\notenumber by 1
            \footnote{$^{\the\notenumber}$}{#1}\tenrm}

\def\ref{\bigskip \centerline{REFERENCES} \medskip}

\def\emet{{\it Econometrica\/ }}
\def\jet{{\it Journal of Economic Theory\/ }}

\def\jme{{\it Journal of Mathematical Economics\/ }}

\def\geb{{\it Games and Economic Behavior\/ }}

\def\jpube{{\it Journal of Public Economics\/ }}

\def\paper#1#2#3#4#5{\noindent\hangindent=20pt#1 (#2), ``#3,'' #4, #5.\par}

\def\book#1#2#3#4{\noindent\hangindent=20pt#1 (#2), {\it #3,} #4.\par}


\bigfont
\def\th{\theta}
\def\Th{\Theta}

{ \ \ }

\vskip 1cm

{\bigtype
\clbf{Dynamic mechanism design: An elementary introduction}
}

\vskip 1cm
\bigskip

\cl{Kiho Yoon}
\cl{Department of Economics, Korea University}
\cl{145 Anam-ro, Seongbuk-gu, Seoul, Korea 02841}
\cl{ \tt kiho@korea.ac.kr}
\cl{\tt http://econ.korea.ac.kr/\~{ }kiho}

\vskip 1cm

\clbf{Abstract}
\medskip

\is{\baselineskip=12pt This paper introduces dynamic mechanism design in an elementary fashion. We first examine optimal dynamic mechanisms: We find necessary and sufficient conditions for perfect Bayesian incentive compatibility and formulate the optimal dynamic mechanism problem. We next examine efficient dynamic mechanisms: We establish the uniqueness of Groves mechanism and investigate budget balance of the dynamic pivot mechanism in some detail for a bilateral trading environment.
This introduction reveals that many results and techniques of static mechanism design can be straightforwardly extended and adapted to the analysis of dynamic settings.}
\smallskip

\is{\baselineskip=12pt Keywords: optimal mechanism, efficient mechanism, Markov process, incentive compatibility, budget balance}

\is{\baselineskip=12pt JEL Classification: C73, D47, D82}

\ve

\Section{INTRODUCTION}

Mechanism design has been very successful both in theory and in applications. Insightful results have been discovered and then applied to the practical tasks of nonlinear pricing, auctions, market design, public good provision, taxation, regulation, etc. While traditional mechanism design literature examines static environments, the research on dynamic mechanism design is flourishing in recent years. Indeed, many real world problems involve long-term relationships over time, and thus dynamic mechanism design would provide new tools as well as implications that the static mechanism design could not offer. There are several excellent surveys on dynamic mechanism design, including Bergemann and Said (2010), Vohra (2012), Bergemann and Pavan (2015), Pavan (2017), and Bergemann and V\"alim\"aki (2019).

The purpose of this paper is to introduce dynamic mechanism design in an elementary fashion. It is elementary since, first of all, it presents simple frameworks to analyze, and secondly and more importantly, it does not require advanced knowledge for the analysis. In particular, we demonstrate that many results and techniques of static mechanism design can be straightforwardly extended and adapted to the analysis of dynamic settings. Hence, readers with some static mechanism design background but little acquaintance with dynamic mechanism design would find this introduction easy to follow.

We study dynamic settings in which players' private information stochastically evolves over time and decisions are made in each period.\note{There is a strand of dynamic mechanism design that studies settings in which the population of players changes over time, but each player's private information does not. We do not cover it.} The mechanism design literature can be classified into two broad categories: The first one is concerned with optimal mechanisms that maximize the principal's revenue, and the second one is concerned with efficient mechanisms that maximize the social welfare. In static mechanism design, the representative work in the first and second category is, respectively, Myerson (1981) and Vickrey (1961).
\ve

In the next section, we examine optimal dynamic mechanisms. We first find necessary and sufficient conditions for perfect Bayesian incentive compatibility and formulate the optimal dynamic mechanism problem. The technique we employ is quite standard in static mechanism design. In Section 3, we examine efficient dynamic mechanisms. It is well-known in static mechanism design that the Groves mechanism is the only outcome efficient and dominant strategy incentive compatible mechanism. We extend this uniqueness result to dynamic settings. In particular, we closely follow the method of proof in Green and Laffont (1977) to highlight our assertion that many results in static mechanism design can be ported to dynamic settings without novel insight and/or apparatus. A special instance of the dynamic Groves mechanism is the dynamic pivot mechanism of Bergemann and V\"alim\"aki (2010), which is a dynamic version of the famous Vickrey-Clarke-Groves (VCG) mechanism. To see how the transition kernel regarding the evolution of private information affects the performance of dynamic mechanisms, we investigate budget balance of the dynamic pivot mechanism in some detail for a bilateral trading environment. Section 4 concludes.

\Section{OPTIMAL DYNAMIC MECHANISMS}

\cl{2.1. THE SETUP}
\medskip

In this section, we examine optimal dynamic mechanisms. We consider a single-player setting without loss of generality.\note{It is straightforward to extend the results to the multi-player setting. We focus on the single-player setting for notational convenience.} Let $t \in \{1,2, \ldots, T \}$ denote a period, where $T$ may be infinite. The player's type in period $t$, which is private information, is $\th_t \in \Th = [\underline \th, \overline \th]$. After $\th_t$ is realized in period $t$, a public action $a_t \in A$ is determined. In addition, let $z_t \in \Re$ be a monetary transfer from the player in period $t$. Given sequences $(\th_1, \ldots, \th_T)$ of types and $(a_1, \ldots, a_T)$ of actions, together with ($z_1, \ldots, z_T)$ of monetary transfers, the player's total payoff is
$$\sum_{t=1}^T \d^{t-1} \bigl( v(\th_t, a_t) - z_t \bigr),$$
where $\d \in [0,1]$ is the discount factor and $v(\cdot)$ is a (one-period) valuation function.\note{We exclude $\d=1$ when $T=\infty$.}  Let $F_1(\th_1)$ denote the distribution of $\th_1$, with $f_1(\th_1)$ being the corresponding density function. Define $\th^t=(\th_1, \ldots, \th_t)$ and $a^t=(a_1, \ldots, a_t)$, and let $F_t(\th_t |\th^{t-1}, a^{t-1})$ denote the conditional distribution of $\th_t$, with $f_t(\th_t |\th^{t-1}, a^{t-1})$ being the corresponding density function. We impose the following {\it Markov property\/} throughout the paper:
$$F_t(\th_t|\th^{t-1}, a^{t-1}) = F_t(\th_t|\th_{t-1}, a_{t-1}),$$
that is, $F_t$ does not depend on $\th_s$ or $a_s$ for $s=1, \ldots, t-2$.\note{We may alternatively impose the Markov assumption as $F_t(\th_t|\th^{t-1}, a^{t-1}) = F_t(\th_t|\th_{t-1}, a^{t-1})$, i.e., $F_t$ does not depend on $\th_s$ but depends on $a_s$ for $s=1, \ldots, t-2$. This alternative assumption does not affect the following results.}

\medskip
\cl{2.2. TWO-PERIOD CASE}
\smallskip

Let us first discuss the two-period case. A dynamic (direct) mechanism is given by $\a_1: \Th \rightarrow A, \t_1: \Th \rightarrow \Re, \a_2: \Th \times A \times \Th \rightarrow A$, and $\t_2: \Th \times A \times \Th \rightarrow \Re$. Thus, $\a_1(\hat \th_1)$ and $\t_1(\hat \th_1)$ are the action chosen and the transfer, respectively, in period 1 when the player's report is $\hat \th_1$, and $\a_2(\hat \th_1, a_1, \hat \th_2)$ and $\t_2(\hat \th_1, a_1, \hat \th_2)$ are the action chosen and the transfer, respectively, in period 2 when the player's report in period 1 is $\hat \th_1$, the action chosen in period 1 is $a_1$, and the player's report in period 2 is $\hat \th_2$. Note that $a_1$ in $\a_2(\cdot)$ and $\t_2(\cdot)$ is $\a_1(\hat \th_1)$ when the mechanism is implemented. The player's strategy is $\s_1: \Th \rightarrow \Th$ and $\s_2:\Th \times \Th \times A \times \Th \rightarrow \Th$. Thus, $\hat \th_1 = \s_1(\th_1)$ is the report in period 1 when the type is $\th_1$, and $\hat \th_2 = \s_2(\th_1, \hat \th_1, a_1, \th_2)$ is the report in period 2 when the type, report, and action in period 1 are $\th_1$, $\hat \th_1$, and $a_1$, respectively, and the type in period 2 is $\th_2$.

Define
$$U_2(\hat \th_2, \th_2 ; \hat \th_1) = v(\th_2, \a_2(\hat \th_1, \a_1(\hat \th_1), \hat \th_2))-\t_2(\hat \th_1, \a_1(\hat \th_1), \hat \th_2).$$
\noindent This is the player's period-2 payoff when the true type is $\th_2$ but the report is $\hat \th_2$ in period 2 and the report is period 1 is $\hat \th_1$. Note that this payoff does not depend on $\th_1$, the true type in period 1. Define with a slight abuse of notation that $U_2(\th_2; \hat \th_1) = U_2(\th_2, \th_2; \hat \th_1)$. Define
$$\eqalign{U_1(\hat \th_1, \th_1) = \ &v(\th_1, \a_1(\hat \th_1))-\t_1(\hat \th_1) \cr
+ & \ \d \int_{\underline \th}^{\overline \th}\Bigl(v(\tilde \th_2, \a_2(\hat \th_1, \a_1(\hat \th_1),\tilde \th_2))-\t_2(\hat \th_1, \a_1(\hat \th_1),\tilde \th_2)\Bigr) dF_2(\tilde \th_2 | \th_1, \a_1(\hat \th_1)).}$$
\noindent Note that $F_2(\tilde \th_2 | \th_1, \a_1(\hat \th_1))$ depends on the true type $\th_1$, and the chosen action (that depends on the report $\hat \th_1$). Define with a slight abuse of notation that $U_1(\th_1) = U_1(\th_1, \th_1)$.

Incentive compatibility is\note{Note that it suffices to consider only the one-shot deviations in $(IC_1)$ by the unimprovability principle.}
$$U_1(\th_1) \geq U_1(\hat \th_1, \th_1) {\rm \ for \ all \ } \th_1 {\rm \ and \ } \hat \th_1, \eqno(IC_1)$$
$$U_2(\th_2; \hat \th_1) \geq U_2(\hat \th_2, \th_2; \hat \th_1) {\rm \ for \ all \ } \th_2, \hat \th_2 {\rm \ and \ } \hat \th_1. \eqno(IC_2)$$
\noindent Note that $(IC_2)$ does not depend on $\th_1$, the true type in period 1. $(IC_2)$ can be written as
$$U_2(\th_2; \hat \th_1) - U_2(\hat \th_2; \hat \th_1) \geq v(\th_2, \a_2(\hat \th_1, \a_1(\hat \th_1), \hat \th_2))-v(\hat \th_2, \a_2(\hat \th_1, \a_1(\hat \th_1), \hat \th_2)).$$
Interchanging the roles of $\th_2$ and $\hat \th_2$, we have
$$U_2(\hat \th_2; \hat \th_1) - U_2(\th_2; \hat \th_1) \geq v(\hat \th_2, \a_2(\hat \th_1, \a_1(\hat \th_1), \th_2))-v(\th_2, \a_2(\hat \th_1, \a_1(\hat \th_1), \th_2)).$$
Combining these inequalities,
$$\eqalign{&v(\hat \th_2, \a_2(\hat \th_1, \a_1(\hat \th_1), \hat \th_2))- v(\th_2, \a_2(\hat \th_1, \a_1(\hat \th_1), \hat \th_2)) \cr
\geq \ &U_2(\hat \th_2; \hat \th_1) - U_2(\th_2; \hat \th_1) \cr
\geq \ &v(\hat \th_2, \a_2(\hat \th_1, \a_1(\hat \th_1), \th_2))-v(\th_2, \a_2(\hat \th_1, \a_1(\hat \th_1), \th_2)).} \eqno(1)$$
With suitable differentiability assumptions, we can get the following formula (2) as well as (6) below. We will assume throughout this section that both $v(\cdot)$ and $f_2(\cdot)$ are continuously differentiable and that both $\a_1(\cdot)$ and $\a_2(\cdot)$ are differentiable almost everywhere. Dividing (1) by $\hat \th_2 - \th_2$ and taking limits, we get
$$\frac{dU_2(\th_2; \hat \th_1)}{d\th_2} = v_\th(\th_2, \a_2(\hat \th_1, \a_1(\hat \th_1), \th_2)) \eqno(2)$$
\noindent almost everywhere. Note that the notation $v_\th(\th,a)$ is the partial derivative of $v(\th,a)$ with respect to $\th$.

We assume that $\partial v(\th,a)/\partial \th \geq 0$ and $\partial^2 v(\th,a)/(\partial\th\partial a) \geq 0$ hold. Note that this is the single-crossing condition. Then, $(1)$ implies the monotonicity property of
$$\a_2(\hat \th_1, \a_1(\hat \th_1), \hat \th_2) \geq \a_2(\hat \th_1, \a_1(\hat \th_1), \th_2) {\rm \ \ \ for \ all \ } \hat \th_2 > \th_2 {\rm \ and \ } \hat \th_1. \eqno(3)$$
\noindent Next, since $dU_2(\th_2; \hat \th_1)/d\th_2$ is continuous almost everywhere on the interval $[\underline \th, \overline \th]$, it is Riemann integrable and we have
$$U_2(\th_2; \hat \th_1) = U_2(\underline \th; \hat \th_1) + \int_{\underline \th}^{\th_2} v_\th(\tilde \th_2, \a_2(\hat \th_1, \a_1(\hat \th_1), \tilde \th_2))d\tilde \th_2 {\rm \ for \  all \ } \th_2 {\rm \ and \ } \hat \th_1. \eqno(4)$$

We have thus far shown that $(IC_2)$ implies (3) and (4). It is easy to show that the converse also holds. Suppose not. Then, there exists $\hat \th_1, \th_2$, and $\hat \th_2$ such that $U_2(\hat \th_2, \th_2; \hat \th_1) > U_2(\th_2; \hat \th_1)$, which implies
$$v(\th_2, \a_2(\hat \th_1, \a_1(\hat \th_1), \hat \th_2))-v(\hat \th_2, \a_2(\hat \th_1, \a_1(\hat \th_1), \hat \th_2)) > U_2(\th_2; \hat \th_1) - U_2(\hat \th_2; \hat \th_1).$$
The left-hand side (LHS henceforth) is
$$\int_{\hat \th_2}^{\th_2} v_\th(\tilde \th_2, \a_2(\hat \th_1, \a_1(\hat \th_1), \hat \th_2)) d\tilde \th_2,$$
and the right-hand side (RHS henceforth) is
$$\int_{\hat \th_2}^{\th_2} v_\th(\tilde \th_2, \a_2(\hat \th_1, \a_1(\hat \th_1), \tilde \th_2))d\tilde \th_2$$
by (4).
Rearranging,
$$\int_{\hat \th_2}^{\th_2} \bigl( v_\th(\tilde \th_2, \a_2(\hat \th_1, \a_1(\hat \th_1), \hat \th_2)) - v_\th(\tilde \th_2, \a_2(\hat \th_1, \a_1(\hat \th_1), \tilde \th_2)) \bigr)d\tilde \th_2 > 0.$$
But, the single-crossing assumption and the monotonicity (3) implies that this is not possible. In summary, we have:

\thm1 $(IC_2)$ holds if and only if $(3)$ and $(4)$ hold. \ok

As for period 1, $(IC_1)$ can be written as
$$\eqalign{U_1(\th_1)-U_1(\hat \th_1) \geq \ &v(\th_1, \a_1(\hat \th_1))-v(\hat \th_1, \a_1(\hat \th_1)) \cr
+ &\ \d \int_{\underline \th}^{\overline \th} U_2(\tilde \th_2; \hat \th_1) d\bigl(F_2(\tilde \th_2|\th_1, \a_1(\hat \th_1)) - F_2(\tilde \th_2|\hat \th_1, \a_1(\hat \th_1))\bigr).}$$
Interchanging the roles of $\th_1$ and $\hat \th_1$, we have
$$\eqalign{U_1(\hat \th_1)-U_1(\th_1) \geq \ &v(\hat \th_1, \a_1(\th_1))-v(\th_1, \a_1(\th_1)) \cr
+ &\ \d \int_{\underline \th}^{\overline \th} U_2(\tilde \th_2;\th_1) d\bigl(F_2(\tilde \th_2|\hat \th_1, \a_1(\th_1)) - F_2(\tilde \th_2|\th_1, \a_1(\th_1))\bigr).}$$
Combining these inequalities,
{\smalltype
$$\eqalign{&v(\hat \th_1, \a_1(\hat \th_1))-v(\th_1, \a_1(\hat \th_1)) + \d \int_{\underline \th}^{\overline \th} U_2(\tilde \th_2; \hat \th_1) d\bigl(F_2(\tilde \th_2|\hat \th_1, \a_1(\hat \th_1)) - F_2(\tilde \th_2|\th_1, \a_1(\hat \th_1))\bigr) \cr
\geq \ &U_1(\hat \th_1) - U_1(\th_1) \cr
\geq \ &v(\hat \th_1, \a_1(\th_1))-v(\th_1, \a_1(\th_1)) + \d \int_{\underline \th}^{\overline \th} U_2(\tilde \th_2;\th_1) d\bigl(F_2(\tilde \th_2|\hat \th_1, \a_1(\th_1)) - F_2(\tilde \th_2|\th_1, \a_1(\th_1))\bigr).}\eqno(5)$$
}
Dividing by $\hat \th_1 - \th_1$ and taking limits, we get
$$\frac{dU_1(\th_1)}{d\th_1} = v_\th(\th_1, \a_1(\th_1)) + \d \int_{\underline \th}^{\overline \th} U_2(\tilde \th_2;\th_1) \ \frac{\partial f_2(\tilde \th_2|\th_1, \a_1(\th_1))}{\partial \th_1} d\tilde \th_2$$
almost everywhere. Note that $\partial f_2/\partial \th_1$ is only with respect to $\th_1$ in $f_2(\th_2|\th_1, a_1)$, not with respect to $a_1$. Now,
$$\eqalign{&\int_{\underline \th}^{\overline \th} U_2(\tilde \th_2;\th_1) \ \frac{\partial f_2(\tilde \th_2|\th_1, \a_1(\th_1))}{\partial \th_1} d\tilde \th_2= \Bigl[U_2(\tilde \th_2; \th_1) \ \frac{\partial F_2(\tilde \th_2|\th_1, \a_1(\th_1))}{\partial \th_1}\Bigr]_{\underline \th}^{\overline \th} \cr
&-\int_{\underline \th}^{\overline \th} v_\th(\tilde \th_2, \a_2(\th_1, \a_1(\th_1), \tilde \th_2)) \ \frac{\partial F_2(\tilde \th_2|\th_1, \a_1(\th_1))}{\partial \th_1} d\tilde\th_2 \cr
= &- \int_{\underline \th}^{\overline \th} v_\th(\tilde \th_2, \a_2(\th_1, \a_1(\th_1), \tilde \th_2)) \ \frac{\partial F_2(\tilde \th_2|\th_1, \a_1(\th_1))}{\partial \th_1} d\tilde\th_2,}$$
where the first equality follows from (2) and the second equality follows from the fact that $F_2(\underline \th | \th_1, \a_1(\th_1))=0$ and $F_2(\overline \th | \th_1, \a_1(\th_1))=1$ for all $\th_1$ and so $\partial F_2/\partial \th_1 = 0$ when $\tilde \th_2=\underline \th$ or $\overline \th$. Therefore,
$$\frac{dU_1(\th_1)}{d\th_1} = v_\th(\th_1, \a_1(\th_1))- \d \int_{\underline \th}^{\overline \th} v_\th(\tilde \th_2, \a_2(\th_1, \a_1(\th_1), \tilde \th_2)) \ \frac{\partial F_2(\tilde \th_2|\th_1, \a_1(\th_1))}{\partial \th_1} d\tilde\th_2 \eqno(6)$$
almost everywhere. We thus have
$$\eqalign{U_1(\th_1)= \ &U_1(\underline \th) + \int_{\underline \th}^{\th_1} v_\th(\tilde \th_1, \a_1(\tilde \th_1)) d\tilde \th_1 \cr
- \ & \d \int_{\underline \th}^{\th_1} \int_{\underline \th}^{\overline \th} v_\th(\tilde \th_2, \a_2(\tilde \th_1, \a_1(\tilde \th_1), \tilde \th_2)) \ \frac{\partial F_2(\tilde \th_2|\tilde \th_1, \a_1(\tilde \th_1))}{\partial \th_1} d\tilde\th_2 d\tilde \th_1} \eqno(7)$$
for all $\th_1$.
\smallskip

We have shown that $(IC_1)$ implies (7). We next show that (7) and the following condition together with (4) imply $(IC_1)$.
{\smalltype

$$\eqalign{&\int_{\hat \th_1}^{\th_1} v_\th(\tilde \th_1, \a_1(\tilde \th_1)) d\tilde \th_1 - \d \int_{\hat \th_1}^{\th_1} \int_{\underline \th}^{\overline \th} v_\th(\tilde \th_2, \a_2(\tilde \th_1, \a_1(\tilde \th_1), \tilde \th_2)) \frac{\partial F_2(\tilde \th_2|\tilde \th_1, \a_1(\tilde \th_1))}{\partial \th_1} d\tilde \th_2 d\tilde \th_1 \cr
\geq & \int_{\hat \th_1}^{\th_1} v_\th(\tilde \th_1, \a_1(\hat \th_1)) d\tilde \th_1 - \d \int_{\hat \th_1}^{\th_1} \int_{\underline \th}^{\overline \th} v_\th(\tilde \th_2, \a_2(\hat \th_1, \a_1(\hat \th_1), \tilde \th_2)) \frac{\partial F_2(\tilde \th_2|\tilde \th_1, \a_1(\hat \th_1))}{\partial \th_1} d\tilde \th_2 d\tilde \th_1} \eqno(8)$$
}
\noindent for all $\th_1$ and $\hat \th_1$. Observe first that, by (7), the LHS of (8) is equal to $U_1(\th_1) - U_1(\hat \th_1)$. Observe next that the RHS is equal to

$$\eqalign{&v(\th_1, \a_1(\hat \th_1))-v(\hat \th_1, \a_1(\hat \th_1)) \cr
- & \d \int_{\hat \th_1}^{\th_1} \int_{\underline \th}^{\overline \th} v_\th(\tilde \th_2, \a_2(\hat \th_1, \a_1(\hat \th_1), \tilde \th_2)) \frac{\partial F_2(\tilde \th_2|\tilde \th_1, \a_1(\hat \th_1))}{\partial \th_1} d\tilde \th_2 d\tilde \th_1 \cr
= \ &v(\th_1, \a_1(\hat \th_1))-v(\hat \th_1, \a_1(\hat \th_1)) \cr
- & \d \int_{\underline \th}^{\overline \th} v_\th(\tilde \th_2, \a_2(\hat \th_1, \a_1(\hat \th_1), \tilde \th_2)) \int_{\hat \th_1}^{\th_1}  \frac{\partial F_2(\tilde \th_2|\tilde \th_1, \a_1(\hat \th_1))}{\partial \th_1} d\tilde \th_1 d\tilde \th_2 \cr
= \ &v(\th_1, \a_1(\hat \th_1))-v(\hat \th_1, \a_1(\hat \th_1)) \cr
- & \d \int_{\underline \th}^{\overline \th} v_\th(\tilde \th_2, \a_2(\hat \th_1, \a_1(\hat \th_1), \tilde \th_2)) \bigl(F_2(\tilde \th_2|\th_1, \a_1(\hat \th_1))-F_2(\tilde \th_2|\hat \th_1, \a_1(\hat \th_1)) \bigr) d\tilde \th_2 }$$
$$\eqalign{= \ &v(\th_1, \a_1(\hat \th_1))-v(\hat \th_1, \a_1(\hat \th_1)) \cr
- & \d \Bigl[ U_2(\tilde \th_2; \hat \th_1) \bigl(F_2(\tilde \th_2|\th_1, \a_1(\hat \th_1))-F_2(\tilde \th_2|\hat \th_1, \a_1(\hat \th_1)) \bigr)\Bigr]_{\underline \th}^{\overline \th} \cr
+ & \d \int_{\underline \th}^{\overline \th} U_2(\tilde \th_2; \hat \th_1) d \bigl(F_2(\tilde \th_2|\th_1, \a_1(\hat \th_1))-F_2(\tilde \th_2|\hat \th_1, \a_1(\hat \th_1)) \bigr) \cr
= \ &v(\th_1, \a_1(\hat \th_1))-v(\hat \th_1, \a_1(\hat \th_1)) \cr
+ & \d \int_{\underline \th}^{\overline \th} U_2(\tilde \th_2; \hat \th_1) d \bigl(F_2(\tilde \th_2|\th_1, \a_1(\hat \th_1))-F_2(\tilde \th_2|\hat \th_1, \a_1(\hat \th_1)) \bigr).}$$

\noindent The first equality follows from the change in the order of integration, the second equality follows from integrating out the inner integral, the third equality follows from (4) and integration by parts, and the last equality follows from the fact that $F_2(\underline \th|\th_1, \a_1(\hat \th_1))=F_2(\underline \th|\hat \th_1, \a_1(\hat \th_1))=0$ and $F_2(\overline \th|\th_1, \a_1(\hat \th_1))=F_2(\overline \th|\hat \th_1, \a_1(\hat \th_1))=1$. Putting together, this is nothing but $(IC_1)$, and we proved the claim. It is straightforward to see that $(IC_1)$ implies (8): Follow the reverse steps of the previous argument. Hence, we have:

\thm2 Assume that (4) holds. Then, $(IC_1)$ holds if and only if (7) and (8) hold. \ok

By definition of $U_1(\th_1)$, the total expected payment the player makes is
$$\eqalign{&\int_{\underline \th}^{\overline \th} v(\tilde \th_1, \a_1(\tilde \th_1)) f_1(\tilde \th_1)d\tilde \th_1 \cr
+ & \d \int_{\underline \th}^{\overline \th} \int_{\underline \th}^{\overline \th} v(\tilde \th_2, \a_2(\tilde \th_1, \a_1(\tilde \th_1), \tilde \th_2)) f_2(\tilde \th_2|\tilde \th_1, \a_1(\tilde \th_1)) f_1(\tilde \th_1) d\tilde \th_2 d\tilde \th_1 \cr
- & \int_{\underline \th}^{\overline \th} U_1(\tilde \th_1) f_1(\tilde \th_1) d\tilde \th_1.}$$
Since
$$\eqalign{&\int_{\underline \th}^{\overline \th} U_1(\tilde \th_1) f_1(\tilde \th_1) d\tilde \th_1 = \Bigl[-U_1(\tilde \th_1) (1-F_1(\tilde \th_1))\Bigr]_{\underline \th}^{\overline \th} + \int_{\underline \th}^{\overline \th} \frac{dU_1(\tilde \th_1)}{d\th_1} (1-F_1(\tilde \th_1))d\tilde \th_1 \cr
= \ & U_1(\underline \th) + \int_{\underline \th}^{\overline \th} v_\th(\tilde \th_1, \a_1(\tilde \th_1))(1-F_1(\tilde \th_1)) d\tilde \th_1 \cr
- \ & \d \int_{\underline \th}^{\overline \th} \int_{\underline \th}^{\overline \th} v_\th(\tilde \th_2, \a_2(\tilde \th_1, \a_1(\tilde \th_1), \tilde \th_2)) \ \frac{\partial F_2(\tilde \th_2|\tilde \th_1, \a_1(\tilde \th_1))}{\partial \th_1} (1-F_1(\tilde \th_1)) d\tilde\th_2 d\tilde\th_1}$$
where the second equality holds by the differential form of $(7)$, the total expected payment is equal to
$$\eqalign{&\int_{\underline \th}^{\overline \th} \Bigl[v(\tilde \th_1, \a_1(\tilde \th_1)) - v_\th(\tilde \th_1, \a_1(\tilde \th_1)) \frac{1-F_1(\tilde \th_1)}{f_1(\tilde \th_1)}\Bigr] f_1(\tilde \th_1) d\tilde \th_1 \cr
+ \ & \d \int_{\underline \th}^{\overline \th} \int_{\underline \th}^{\overline \th} \Bigl[ v(\tilde \th_2, \a_2(\tilde \th_1, \a_1(\tilde \th_1), \tilde \th_2))+ v_\th(\tilde \th_2, \a_2(\tilde \th_1, \a_1(\tilde \th_1), \tilde \th_2)) \cr
& \ \ \ \times \frac{1-F_1(\tilde \th_1)}{f_1(\tilde \th_1)}\frac{\partial F_2(\tilde \th_2|\tilde \th_1, \a_1(\tilde \th_1))/\partial \th_1}{f_2(\tilde \th_2|\tilde \th_1, \a_1(\tilde \th_1))}\Bigr]f_2(\tilde \th_2|\tilde \th_1, \a_1(\tilde \th_1))f_1(\tilde \th_1)d\tilde \th_2 d\tilde \th_1 \cr
- \ & U_1(\underline \th).}$$

Let us specialize to the situation where a monopolistic seller wants to sell an indivisible good to a potential buyer. Then, the buyer's payoff is $v(\th,a)=\th a = \th q$, where $q$ is the probability of trade. In this case, we have $v(\th_1, \a_1(\th_1))=\th_1 q_1(\th_1)$ and $v(\th_2, \a_2(\th_1, \a_1(\th_1), \th_2))=\th_2 q_2(\th_1, q_1(\th_1), \th_2)$. Hence, \par
\cl{$v_\th(\th_1, \a_1(\th_1))=q_1(\th_1)$ and $v_\th(\th_2, \a_2(\th_1, \a_1(\th_1), \th_2))=q_2(\th_1, q_1(\th_1), \th_2)$.}
\noindent The seller's revenue is\note{Note that $\a_1(\tilde \th_1) = q_1(\tilde \th_1)$ and $\a_2(\tilde \th_1, \a_1(\tilde \th_1), \tilde \th_2) = q_2(\tilde \th_1, q_1(\tilde \th_1), \tilde \th_2)$.}
$$\eqalign{&\int_{\underline \th}^{\overline \th} \Bigl[\tilde \th_1 - \frac{1-F_1(\tilde \th_1)}{f_1(\tilde \th_1)}\Bigr] q_1(\tilde \th_1)f_1(\tilde \th_1) d\tilde \th_1 \cr
+ \ & \d \int_{\underline \th}^{\overline \th} \int_{\underline \th}^{\overline \th} \Bigl[ \tilde \th_2 + \frac{1-F_1(\tilde \th_1)}{f_1(\tilde \th_1)}\frac{\partial F_2(\tilde \th_2|\tilde \th_1, q_1(\tilde \th_1))/\partial \th_1}{f_2(\tilde \th_2|\tilde \th_1, q_1(\tilde \th_1))}\Bigr] q_2(\tilde \th_1, q_1(\tilde \th_1), \tilde \th_2) \cr
& \ \ \ \times  f_2(\tilde \th_2|\tilde \th_1, q_1(\tilde \th_1))f_1(\tilde \th_1)d\tilde \th_2 d\tilde \th_1 \cr
- \ & U_1(\underline \th).}$$
If we define
$$\psi_1(\th_1) = \th_1 - \frac{1-F_1(\th_1)}{f_1(\th_1)};$$
$$\psi_2(\th_1, \th_2) = \th_2 + \frac{1-F_1(\th_1)}{f_1(\th_1)}\frac{\partial F_2(\th_2|\th_1, q_1(\th_1))/\partial \th_1}{f_2(\th_2| \th_1, q_1(\th_1))},$$
then the seller's revenue becomes
$$\eqalign{&\int_{\underline \th}^{\overline \th} \psi_1(\tilde \th_1) q_1(\tilde \th_1)f_1(\tilde \th_1) d\tilde \th_1 \cr
+ \ & \d \int_{\underline \th}^{\overline \th} \int_{\underline \th}^{\overline \th}  \psi_2(\tilde \th_1, \tilde \th_2) q_2(\tilde \th_1, q_1(\tilde \th_1), \tilde \th_2) f_2(\tilde \th_2|\tilde \th_1, q_1(\tilde \th_1))f_1(\tilde \th_1)d\tilde \th_2 d\tilde \th_1 - U_1(\underline \th).}$$

\noindent Observe that $\psi_1(\th_1)$ and $\psi_2(\th_1, \th_2)$ correspond to the virtual valuation of Myerson (1981). In particular, the term $-\frac{\partial F_2/\partial \th_1}{f_2}$ in $\psi_2(\th_1, \th_2)$ measures the effect of $\th_1$ on $\th_2$, and is called a measure of informativeness by Baron and Besanko (1984) and the impulse response by Pavan {\it et al.\/} (2014). Observe also that the seller's revenue does not depend on the transfer rule and thus the revenue equivalence principle applies. The seller's problem is then to choose the decision rules $q_1(\cdot)$ and $q_2(\cdot)$ to maximize the revenue subject to $U_1(\underline \th) \geq 0$, $(3)$, and $(8)$. We will not analyze the seller's problem further in this elementary introduction, but only note that the optimal solution can be found similarly to the static case when $\psi_1(\th_1)$ is increasing in $\th_1$ and $\psi_2(\th_1, \th_2)$ is increasing in both $\th_1$ and $\th_2$.

We end this section by noting that the analysis above can be extended to the general $T$-period case. See also Baron and Besanko (1984) and Pavan {\it et al.\/} (2014) among others for related derivations.

\Section{EFFICIENT DYNAMIC MECHANISMS}

\medskip
\cl{3.1. THE SETUP}
\smallskip

In this section, we examine efficient dynamic mechanisms. There is a set $I = \{1,\ldots ,n\}$ of players and a countable number of periods, indexed by  $t \in \{0,1,\ldots\}$. Player $i$'s type in period $t$ is $\th_i^t \in \Th_i$. We assume that this is private information. Let $\th^t = (\th_1^t, \ldots, \th_n^t)$ and $\Th = \prod_{i =1}^n \Th_i$. We assume that $\Th$ is a Borel space, i.e., a Borel subset of a complete and separable metric space. Let ${\cal B}(\Th)$ be the Borel $\sigma$-algebra on $\Th$. After $\th^t \in \Th$ is realized in period $t$, a public action $a^t \in A$ is determined. We assume that $A$ is a Borel space, with the Borel $\sigma$-algebra ${\cal B} (A)$.\note{We impose the assumption that $\Th$ and $A$ are Borel spaces to employ some of the results in Hern\'andez-Lerma and Lasserre (1996). See footnote 10.} In addition, let $z_i^t \in \Re$ be a monetary transfer from player $i$ in period $t$. Given sequences $(\th^0, \th^1, \ldots)$ of type profiles and $(a^0, a^1, \ldots)$ of actions, together with $(z_i^0, z_i^1, \ldots)$ of $i$'s monetary transfers, player $i$'s total payoff is
$$\sum_{t=0}^{\infty} \d^t \Bigl(v_i(\th_i^t,a^t)-z_i^t\Bigr),$$
where (i) $\d$ is a common discount factor and $\d < 1$, and (ii) $v_i(\cdot)$ is a measurable (one-period) valuation function. The valuation function is usually called as the reward function in the Markov decision process literature. Note that we deal with the private-values environment in that player $i$'s valuation function depends only on player $i$'s type. We assume that $v_i(\cdot)$ is bounded, that is, $|v_i(\th_i, a)| \leq C < \infty$ for all $\th_i$ and $a$.

The dynamic evolution of players' types is represented by a stochastic kernel. Let $p(B | \th^t, a^t)$ for $B \in {\cal B} (\Th)$ be the conditional probability that the type profile lies in $B$ in period $t+1$ when the type profile is $\th^t$ and the action is $a^t$ in period $t$. We have (i) $p(\cdot| \th^t, a^t)$ is a probability measure on $\Th$ for each fixed $(\th^t, a^t)$, and (ii) $p(B | \cdot, \cdot)$ is a measurable function with respect to the product $\sigma$-algebra ${\cal B} (\Th \times A)$ for each fixed $B \in {\cal B}(\Th)$. We assume that $p(\cdot | \cdot, \cdot)$ is independent across players in the sense that $p(\th'|\th,a)= \prod_{i=1}^n p_i(\th'_i|\th_i,a)$. Observe that, except for the fact that $\th$ is private information, this environment fits into a Markov decision process with $\Th$ being the set of states.

We focus attention on dynamic direct mechanisms that ask each player to report his type (i.e., state) in each period. In particular, we will restrict attention to deterministic Markovian mechanisms. A deterministic Markovian decision rule is a measurable function $\hat a^t: \Th \rightarrow A$ that chooses an action based only on current state.\note{A general decision rule may depend on all past reports and actions. It may be deterministic or probabilistic.} In addition, the mechanism specifies the monetary transfers: A deterministic Markovian transfer rule of the mechanism in period $t$ is a collection of measurable functions $\{\hat z_i^t: \Th \rightarrow \Re\}_{i \in I}$. Let $\hat z^t = (\hat z_1^t, \ldots, \hat z_n^t)$. A dynamic direct mechanism is represented by a family of decision rules and monetary transfer rules, $\{\hat a^t, \hat z^t\}_{t=0}^\infty$.

A {\it policy\/} of the mechanism is a sequence of decision rules, that is, a policy is $\pi = (\hat a^0, \hat a^1, \ldots)$. We call a policy {\it stationary\/} if $\hat a^t = \hat a$ for all $t$. A stationary policy has the form $\pi=(\hat a,\hat a,\ldots)$, which is denoted by $\hat a^\infty$. For the stationary environment considered in this paper,\note{The environment is stationary since both the valuation function $v_i(\cdot)$ for all $i$ and the stochastic kernel $p(\cdot | \cdot)$ do not vary with $t$.} we can without loss of generality restrict our attention to deterministic stationary policies when finding a policy that maximizes the expected discounted sum of players' valuations
$$E_\th^\pi \Bigl[ \sum_{t=0}^\infty \d^{t} \sum_{j=1}^n v_j(\tilde \th_j^t, \tilde a^t) \Bigr]$$
for every $\th \in \Th$.\note{See Theorem 4.2.3 of Hern\'andez-Lerma and Lasserre (1996). Note that a deterministic stationary policy is a deterministic Markovian policy.} Note that the expectation is over the stochastic process given the initial $\th$.\note{We will assume throughout that the relevant maximum is attained without specifying sufficient conditions. This assumption is valid under standard conditions on the environment: See Theorem 4.2.3 of Hern\'andez-Lerma and Lasserre (1996) and the discussion preceding it.} An outcome efficient policy thus has the form $\pi^* = (a^*)^\infty$ where $a^*: \Th \rightarrow A$. We can also restrict our attention to stationary transfer rules. We want to note that some previous works in the literature consider only deterministic Markovian mechanisms from the outset without proper theoretical underpinnings, that is, without providing conditions that rationalize this restriction for the particular settings.

\medskip
\cl{3.2. THE UNIQUENESS OF DYNAMIC GROVES MECHANISMS}
\smallskip

Define the total social welfare function $W:\Th \rightarrow \Re$ recursively by the following optimality equation (or Bellman equation):
$$W(\th)=\sum_{j=1}^n v_j(\th_j,a^*(\th))+\d \int_{\Th} W(\th') p(d\th'|\th,a^*(\th)).$$
Given an outcome efficient policy $\pi^*$, we can also define player $i$'s total valuation function $V_i(\th)$  recursively as
$$V_i(\th) = v_i(\th_i, a^*(\th)) + \d \int_{\Th}  V_i(\th') p(d\th'|\th,a^*(\th)).$$
Observe that
$$\eqalign{V_i(\th)=&v_i(\th_i, a^*(\th)) + \d \int_{\Th} v_i(\th'_i, a^*(\th')) p(d\th'|\th,a^*(\th))  \cr
+& \d^2 \int_{\Th} \int_{\Th} v_i(\th''_i, a^*(\th''))  p(d\th''|\th',a^*(\th')) p(d\th'|\th,a^*(\th)) + \cdots .}$$
Likewise, we can define the total valuation function of players other than $i$ recursively as
$$V_{-i}(\th) = \sum_{j \ne i} v_j(\th_j, a^*(\th)) + \d \int_{\Th} V_{-i}(\th') p(d\th'|\th,a^*(\th)).$$

\noindent Note that we use the usual notational convention that the subscript $-i$ pertains to players other than $i$. Thus, $\th_{-i} = (\th_1, \ldots, \th_{i-1}, \th_{i+1}, \ldots,\th_n)$, $\Th_{-i} = \prod_{j \ne i} \Th_j$, and so on. We now define dynamic Groves mechanisms.

\defn1 A dynamic Groves mechanism is a dynamic direct mechanism with an outcome efficient policy  $\pi^* = (a^*)^\infty$ and a stationary total transfer rule for player $i=1, \ldots, n$ given as
$$Z^*_i(\th) = - V_{-i}(\th) + \Phi_i(\th_{-i})$$
for some $\Phi_i: \Th_{-i} \rightarrow \Re$. \ok

Note that $\Phi_i(\cdot)$ does not depend on $\th_i$. If we recall the terminology of d'Aspremont and G\'erard-Varet (1979), the dynamic Groves mechanism is a {\it distribution mechanism\/} since the total transfer rule is given as the difference between $V_{-i}(\th)$ and the total distribution rule $\Phi_i(\th_{-i})$. In addition, the total distribution rule $\Phi_i(\th_{-i})$ is {\it discretionary\/} because it does not depend on $\th_i$.

It is easy to establish that dynamic Groves mechanisms are periodic ex-post incentive compatible, that is, the truth-telling strategy is a best response for every player $i$ and every true type profile $\th$ in every period $t$ and private history.\note{For a more detailed discussion on the concept of ex-post incentive compatibility in dynamic settings, see Bergemann and V\"alim\"aki (2010), Yoon (2021), etc.}

\thm3 A dynamic Groves mechanism is periodic ex-post incentive compatible. \ok

\pf Omitted since it is straightforward. See, for instance, Yoon (2021). \endpf

We now establish the uniqueness of dynamic Groves mechanisms. Our approach is to port the results for static Groves mechanisms to the dynamic setting: We closely follow the method of proof in Green and Laffont (1977) to highlight our approach of porting the results for static Groves mechanisms to the dynamic setting. Cavallo (2008) has done essentially the same analysis. Hence, the material in this subsection may be taken as a (hopefully) clearer derivation with solid groundwork.

A key step is to define player $i$'s total valuation when the current-period type profile is $(\th_i, \th_{-i})$, the action $a$ is chosen in the current period, and the outcome efficient policy is followed afterwards. Let
$$\eqalign{V_i^O(\th_i, \th_{-i}, a)= & \ v_i(\th_i, a) + \d \int_{\Th} v_i(\th'_i, a^*(\th')) p(d\th'|\th_i, \th_{-i}, a) \cr
+ &\ \d^2 \int_{\Th} \int_{\Th} v_i(\th''_i, a^*(\th'')) p(d\th''|\th',a^*(\th')) p(d\th'|\th_i, \th_{-i}, a) + \cdots .}$$
In recursive form, we have
$$V_i^O(\th_i, \th_{-i}, a) = v_i(\th_i, a) + \d \int_{\Th} V_i^O(\th'_i, \th'_{-i}, a^*(\th')) p(d\th'|\th_i, \th_{-i}, a).$$
Note that player $i$'s total valuation function $V_i(\th)$ defined earlier is equal to $V_i^O(\th, a^*(\th))$. We can similarly define $V_{-i}^O(\th_i, \th_{-i}, a)$ and $W^O(\th_i, \th_{-i}, a)$. We also have $V_{-i}(\th)=V_{-i}^O(\th,$ $ a^*(\th))$ and $W(\th)=W^O(\th, a^*(\th))$. We have:

\thm4 If a dynamic direct mechanism with an outcome efficient policy $\pi^* = (a^*)^\infty$ is periodic ex-post incentive compatible, then it is a dynamic Groves mechanism. \ok

It is convenient to present the following definition and lemma before the proof of this theorem.

\defn2 A dynamic direct mechanism with an outcome efficient policy $\pi^* = (a^*)^\infty$ and a stationary total transfer rule $Z_i: \Th \rightarrow \Re$ satisfies Property A if \par
\cl{$Z_i(\th_i, \th_{-i}) - Z_i(\bar \th_i, \th_{-i}) = V_{-i}(\bar \th_i, \th_{-i}) - V_{-i}(\th_i, \th_{-i})$}
\noindent for all $\th_i, \bar \th_i$, and $\th_{-i}$. \ok

\lemma1 A dynamic direct mechanism with an outcome efficient policy $\pi^* = (a^*)^\infty$ is a dynamic Groves mechanism if and only if it satisfies Property A. \ok

\pf It is obvious that a dynamic Groves mechanism satisfies Property A. For the other direction, define $\Phi_i(\th) = Z_i(\th)+ V_{-i}(\th)$ for the given mechanism. Note that $\Phi_i(\cdot)$ does not depend on $\th_i$, i.e., $\Phi_i(\th_i, \th_{-i}) = \Phi_i(\bar \th_i, \th_{-i})$ by Property A, so write it as $\Phi_i(\th_{-i})$. Then, the total transfer rule given as $Z_i(\th) = -V_{-i}(\th)+\Phi_i(\th_{-i})$ constitutes a dynamic Groves mechanism. \endpf

\pfo{Theorem 4} We will show that if a dynamic direct mechanism with an outcome efficient policy $\pi^* = (a^*)^\infty$ is periodic ex-post incentive compatible then it satisfies Property A. Then, Lemma 1 gives the desired result.

We first establish that, if $a^*(\th_i, \th_{-i}) = a^*(\bar \th_i, \th_{-i})$ and $p(B | \th_i, \th_{-i}, a^*(\bar \th_i, \th_{-i})) = $ $p(B | \bar \th_i, \th_{-i}, a^*(\bar \th_i, \th_{-i}))$ for all $B \in {\cal B}(\Th)$, then $Z_i(\th_i, \th_{-i}) = Z_i(\bar \th_i, \th_{-i})$. Suppose otherwise. Then, there exist $\th_i, \bar \th_i, \th_{-i}$ with $a^*(\th_i, \th_{-i}) = a^*(\bar \th_i, \th_{-i})$ and $p(B | \th_i, \th_{-i}, a^*(\bar \th_i, \th_{-i}))=$ $p(B | \bar \th_i, \th_{-i}, a^*(\bar \th_i, \th_{-i}))$ for all $B \in {\cal B}(\Th)$ but $Z_i(\th_i, \th_{-i}) > Z_i(\bar \th_i, \th_{-i})$.

Now if player $i$ reports $\bar \th_i$ when his true type is $\th_i$, his total payoff is
$$\eqalign{&v_i(\th_i, a^*(\bar \th_i, \th_{-i}))+\d \int_{\Th} V_i(\th') p(d\th' | \th_i, \th_{-i}, a^*(\bar \th_i, \th_{-i})) \cr
&-z_i(\bar \th_i, \th_{-i})-\d \int_{\Th} Z_i(\th') p(d\th' | \th_i, \th_{-i}, a^*(\bar \th_i, \th_{-i})).}$$
Observe that the first two terms are equal to $V_i(\th_i, \th_{-i})$ since $a^*(\th_i, \th_{-i}) = a^*(\bar \th_i, \th_{-i})$ and the next two terms are equal to $-Z_i(\bar \th_i, \th_{-i})$ since $p(d\th'| \th_i, \th_{-i}, a^*(\bar \th_i, \th_{-i})) = p(d\th' | \bar \th_i, \th_{-i},$ $ a^*(\bar \th_i, \th_{-i}))$. Thus, player $i$ has an incentive to report $\bar \th_i$ when his true type is $\th_i$ since
$$V_i(\th) - Z_i(\th) < V_i(\th) - Z_i(\bar \th_i,  \th_{-i}).$$
This contradicts the fact that the mechanism is periodic ex-post incentive compatible.

Suppose next that Property A does not hold. Then, there exist $\th_i, \bar \th_i, \th_{-i}$ with $Z_i(\th_i, \th_{-i}) - Z_i(\bar \th_i, \th_{-i}) = V_{-i}(\bar \th_i, \th_{-i}) - V_{-i}(\th_i, \th_{-i}) - \e$ for some $\e > 0$. Let $\hat \th_i$ be such that

$$\eqalign{{\rm (i) \ \ } & V_i^O(\hat \th_i, \th_{-i}, a^*(\th_i, \th_{-i})) = - V_{-i}(\th_i, \th_{-i}) {\rm \ and \ } \cr
&p(B | \hat \th_i, \th_{-i}, a^*(\th_i, \th_{-i})) = p(B | \th_i, \th_{-i}, a^*(\th_i, \th_{-i})) {\rm \ \ for \  all \ \ } B \in {\cal B}(\Th), \cr
{\rm (ii) \ \ } & V_i^O(\hat \th_i, \th_{-i}, a^*(\bar \th_i, \th_{-i})) =  - V_{-i}(\bar \th_i, \th_{-i}) + \eta  {\rm \ \ with \ \ } 0  < \eta < \e {\rm \ and \ } \cr
&p(B | \hat \th_i, \th_{-i}, a^*(\bar \th_i, \th_{-i})) = p(B | \bar \th_i, \th_{-i}, a^*(\bar \th_i, \th_{-i})) {\rm \ \ for \  all \ \ } B \in {\cal B}(\Th), \cr
{\rm (iii) \ \ } & V_i^O(\hat \th_i, \th_{-i}, a) = -c {\rm \ \  for \ all \ } a \ne a^*(\th_i, \th_{-i}) {\rm \ or \ } a^*(\bar \th_i, \th_{-i}) \cr
&{\rm \  with  \ } c > \sup_{\pi, \th} E_{\th}^{\pi} \Bigl[ \sum_{t=0}^\infty \d^{t} \sum_{j \ne i} v_j(\tilde \th_j^t, \tilde a^t) \Bigr].}$$
We have $a^*(\hat \th_i, \th_{-i}) = a^*(\bar \th_i, \th_{-i})$, that is, $W(\hat \th_i, \th_{-i})$ is maximized at $a^*(\bar \th_i, \th_{-i})$. To see this, observe that, when the current-period type profile is $(\hat \th_i, \th_{-i})$, the action $a^*(\bar \th_i, \th_{-i})$ gives
$$\eqalign{&v_i(\hat \th_i, a^*(\bar \th_i, \th_{-i})) + \sum_{j \ne i} v_j(\th_j, a^*(\bar \th_i, \th_{-i})) + \d \int_{\Th} W(\th') p(d\th' | \hat \th_i, \th_{-i}, a^*(\bar \th_i, \th_{-i})) \cr
= & V_i^O(\hat \th_i, \th_{-i}, a^*(\bar \th_i, \th_{-i})) + V_{-i}(\bar \th_i, \th_{-i}) = \eta}$$
since $p(B | \hat \th_i, \th_{-i}, a^*(\bar \th_i, \th_{-i})) = p(B | \bar \th_i, \th_{-i}, a^*(\bar \th_i, \th_{-i}))$ for all $B \in {\cal B}(\Th)$. Thus, the sum of players' total valuations is equal to $\eta$. Likewise, the action $a^*(\th_i, \th_{-i})$ gives the sum of players' total valuations as zero, and any other action $a$ gives the sum of players' total valuations as less than zero. Hence, $a^*(\hat \th_i, \th_{-i}) = a^*(\bar \th_i, \th_{-i})$. This, together with $p(B | \hat \th_i, \th_{-i}, a^*(\bar \th_i, \th_{-i})) = p(B | \bar \th_i, \th_{-i}, a^*(\bar \th_i, \th_{-i}))$ for all $B \in {\cal B}(\Th)$,  in turn implies that $Z_i(\hat \th_i, \th_{-i}) = Z_i(\bar \th_i, \th_{-i})$ by the first part of the proof.

Since
$$\eqalign{&Z_i(\th_i, \th_{-i}) - Z_i(\bar \th_i, \th_{-i}) = V_{-i}(\bar \th_i, \th_{-i}) - V_{-i}(\th_i, \th_{-i})-\e \cr
= & V_i^O(\hat \th_i, \th_{-i}, a^*(\th_i, \th_{-i})) -  V_i^O(\hat \th_i, \th_{-i}, a^*(\bar \th_i, \th_{-i}))-\e+\eta,}$$
we get
$$V_i^O(\hat \th_i, \th_{-i}, a^*(\th_i, \th_{-i})) - Z_i(\th_i, \th_{-i}) > V_i^O(\hat \th_i, \th_{-i}, a^*(\bar \th_i, \th_{-i})) - Z_i(\bar \th_i, \th_{-i}).$$
Thus, player $i$ has an incentive to report $\th_i$ when his true type is $\hat \th_i$. This contradicts the fact that the mechanism is periodic ex-post incentive compatible.
\endpf

By Theorems 3 and 4, a dynamic direct mechanism with an outcome efficient policy $\pi^* = (a^*)^\infty$ is periodic ex-post incentive compatible if and only if it is a dynamic Groves mechanism. We note that this result is obtained for unrestricted domain in the sense that, as the proof shows, any total valuation $V_i^O(\hat \th_i, \th_{-i}, a)$ and transition kernel $p(\cdot|\hat \th_i, \th_{-i}, a)$ may be constructed as needed. As a matter of fact, the uniqueness result can be established as well on more restricted domains, such as the domain of continuous (or connected, concave, etc.) total valuations, by appropriately porting the corresponding results, say Walker (1978) or Holmstr\"om (1979), for static mechanism design. See Yoon (2021) for an example of this approach, which builds on the more recent work of Carbajal (2010).

\medskip
\cl{3.3. BUDGET BALANCE OF DYNAMIC PIVOT MECHANISMS}
\smallskip

A special instance of the dynamic Groves mechanism is the dynamic pivot mechanism as defined by Bergemann and V\"alim\"aki (2010): Set the function $\Phi_i(\th_{-i})$ in Definition 1 to be equal to
$$W_{-i}(\th_{-i}) = \sum_{j \ne i} v_j(\th_j, a_{-i}^*(\th_{-i})) + \d \int_{\Th_{-i}} W_{-i}(\th'_{-i}) p_{-i}(d\th'_{-i}|\th_{-i},a_{-i}^*(\th_{-i})),$$
where $a_{-i}^*: \Th_{-i} \rightarrow A$ is a decision rule that maximizes the expected discounted sum $E_\th^\pi [ \sum_{t=0}^\infty \d^{t} \sum_{j \ne i} v_j(\tilde \th_j^t, \tilde a^t)] $ of the valuations of players other than $i$. Then, player $i$'s total payoff is equal to his total marginal contribution $W(\th) - W_{-i}(\th_{-i})$. Observe that this is the dynamic version of the famous Vickrey-Clarke-Groves (VCG) mechanism. We investigate the budget balance problem of this mechanism. To get a firm grasp of the subject, we will analyze the bilateral trading environment in some detail.

A seller and a buyer have an opportunity to trade in periods $t=0,1, 2 \ldots$, where the seller is endowed with one indivisible unit of a perishable good at the beginning of each period. Let $\th_i^t \in \Th_i$ be player $i$'s valuation for the good in period $t$, where $i=s$ for the seller and $i=b$ for the buyer. The valuations are private information. Note that this is a dynamic version of the bilateral trading under incomplete information, the static version of which was pioneered by Chatterjee and Samuelson (1983) and Myerson and Satterthwaite (1983). Let $\th^t = (\th_s^t, \th_b^t)$ and $\Th = \Th_s \times \Th_b$. After $\th^t \in \Th$ is realized in period $t$, a trading decision $a^t \in A \subseteq [0,1]$ is determined. Here, $a^t$ is the probability of trade, i.e., the probability that the seller hands over the good to the buyer. In addition, let $z_i^t \in \Re$ be a monetary transfer from player $i$ in period $t$.

The dynamic pivot mechanism in this environment is as follows. First, the decision rule is efficient: An efficient decision rule in each period is $a^*: \Th \rightarrow A$ such that
$$a^*(\th_s^t, \th_b^t)=\cases{1 &if $\th_s^t < \th_b^t$\ , \cr
                               0 & otherwise.}$$

\noindent Thus, the seller's payoff from the decision in period $t$, i.e., $v_s(\th_s^t,a^*)$, is zero when $a^*=1$ and $\th_s^t$ when $a^*=0$.  On the other hand, the buyer's payoff from the decision in period $t$, i.e., $v_b(\th_b^t,a^*)$, is $\th_b^t$ when $a^*=1$ and zero when $a^*=0$. Henceforth, we will normalize players' payoffs from autarky to zero. This in particular implies that the seller's payoff from the decision becomes $-\th_s^t$ when $a^*=1$ and zero when $a^*=0$, whereas the buyer's payoff from the decision remains the same.\note{One may envision that the seller actually produces the good with a cost of $\th_s^t$ only after the decision rule dictates the trade.} That is, (i) When $a^* =1$, we have $v_s=-\th_s^t$ and $v_b=\th_b^t$, and (ii) When $a^*=0$, we have $v_s=v_b=0$. Next, the transfer payment $z_i^*(\th^t)$ from the players is such that $z_s^*(\th)=-\th_b^t$ and $z_b^*(\th)=\th_s^t$ when $a^*=1$, and $z_s^*(\th^t)=z_b^*(\th^t)=0$ when $a^*=0$. Indeed, since the seller cannot trade without the buyer and vice versa, the social welfare without one player is always zero. Thus, the transfer rule of the dynamic pivot mechanism becomes \par
\cl{$z_s^*(\th^t)=-v_b(\th_b^t, a^*(\th^t))$ and $z_b^*(\th^t)=-v_s(\th_s^t,a^*(\th^t))$.}

The dynamic pivot mechanism is periodic ex-post incentive compatible. Observe that both players' payoffs are  $\th_b^t - \th_s^t$ when the trade occurs, and zero when the trade does not occur. Hence, each player's payoff in each period is non-negative, so the periodic ex-post participation constraints are satisfied.

The flow budget deficit of the dynamic pivot mechanism is $-z_s^*(\th^t)-z_b^*(\th^t)$, which is equal to $\th_b^t - \th_s^t$ when $\th_s^t < \th_b^t$ and zero otherwise. Therefore, the dynamic pivot mechanism runs a budget deficit even in expectation. To cope with the budget problem, we modify the dynamic pivot mechanism in a way that lump-sum (participation) fees are collected from the players. In a similar spirit, Yoon (2001, 2008) studied the participatory Vickrey-Clarke-Groves mechanism in various static settings.

\medskip
\noindent {\sl 3.3.1. A two-period example}
\smallskip

We first study the case when there are two periods, $t=0,1$. Equivalently, we assume that the seller is endowed with the good only in periods 0 and 1. We assume $\d=1$ for this two-period example to avoid unnecessary complications.

\medskip
\noindent {A. The continuous case}
\smallskip
\noindent {(1) Independent valuations}
\smallskip

Let us assume that $\th_i^t$'s are independently and identically distributed according to the uniform distribution on $[0,1]$ for all $t=0,1$ and $i=s,b$. Thus, $\th_i^t$'s are independent across periods as well as across players. Then, we have
$$E[z_s^1] = -\int_{0}^{1} \int_{\th_s}^{1} \th_b d\th_b d\th_s=-\frac{1}{3}$$
and
$$E[z_b^1]= \int_{0}^{1} \int_{0}^{\th_b} \th_s d\th_s d\th_b = \frac{1}{6}.$$
\noindent Hence, the mechanism runs an expected deficit of $1/6$  in period 1. It is also clear that the mechanism runs an expected deficit of $1/6$ in period 0, too.\note{Note well that we have to take expectation over all possible valuations since the mechanism does not know the players' private information.}

At the beginning of period 0 when player $i$ knows his valuation $\th_i^0$ but not $\th_i^1$, the latter is a random variable. Thus, both players' expected period-1 payoffs are
$$\int_{0}^{1} \int_{\th_s}^{1} (\th_b - \th_s) d\th_b d\th_s = \frac{1}{6}.$$
\noindent So, the total expected payoff of the seller with valuation $\th_s^0$ at the beginning of period 0 is
$$\int_{\th_s^0}^{1} (\th_b^0 - \th_s^0) d\th_b^0 + \frac{1}{6} = \frac{(1-\th_s^0)^2}{2}+\frac{1}{6}$$
and the total expected payoff of the buyer with valuation $\th_b^0$ at the beginning of period 0 is
$$\int_{0}^{\th_b^0} (\th_b^0 - \th_s^0) d\th_s^0 + \frac{1}{6} = \frac{(\th_b^0)^2}{2} + \frac{1}{6} \ .$$

\noindent This gives us the conclusion that, by charging each player a lump-sum fee of $1/6$, (i) the mechanism can make up for the expected deficit of  both period 0 and period 1, and (ii) both players participate in period 0. Therefore, the dynamic pivot mechanism with lump-sum fees achieves efficiency, (ex-ante) budget balance, and individual rationality.

\medskip
\noindent {(2) Persistent valuations}
\smallskip

Let us assume now that $\th_i^0=\th_i^1=\th_i$ for $i=s,b$. That is, each player's valuation is persistent over time. Assume also that $\th_s$ and $\th_b$ are independently and identically distributed according to the uniform distribution on $[0,1]$. Hence, valuations are independent across players but perfectly correlated across periods.

In this case, the budget deficit problem is not alleviated but exacerbated since players know their period-1 valuations at the beginning of period 0. In fact, we essentially face a static problem duplicated. The total expected payoff of the seller with valuation $\th_s$ and of the buyer with valuation $\th_b$ at the beginning of period 0 is $(1-\th_s)^2$ and $\th_b^2$, respectively. To satisfy the participation constraints (specifically for the seller with $\th_s=1$ and the buyer with $\th_b=0$), the mechanism cannot charge any additional fee, and consequently the mechanism runs an expected deficit of $1/3$.

This example is meant to demonstrate that the dependence of valuations across periods is crucial for the budget balance of the dynamic mechanism. The mechanism is ex-ante budget-balancing when valuations are independent across periods. By contrast, the mechanism runs budget deficit when valuations are perfectly correlated across periods. The natural question is: What is the scope of dependence that ensures budget balance?

\medskip
\noindent {B. The discrete case}
\smallskip

To answer this question, let us assume that $\th_i^t \in \{0,1\}$. That is, each player's valuation takes either zero or one. Then, the trade occurs only when $\th_s^t=0$ and $\th_b^t=1$ in an efficient decision rule. So, $z_s^t=-1$ and $z_b^t=0$ when $a^*=1$, and $z_s^t=z_b^t=0$ when $a^*=0$. Both players' payoffs are 1 when the trade occurs and 0 when the trade does not occur.

Assume that the initial distribution of $\th_i^0$ for $i=s,b$ is such that $\th_i^0 = 0$ or $1$ with equal probability of $1/2$. The transition matrix for the seller is given as
$$P_s = \pmatrix{s_{00} & s_{01} \cr
               s_{10} & s_{11} \cr},$$
where $s_{ij}$ for $i,j=0,1$ is the probability that $\th_s^1 = j$ given $\th_s^0=i$. The transition matrix for the buyer is similarly given as
$$P_b = \pmatrix{b_{00} & b_{01} \cr
               b_{10} & b_{11} \cr}.$$

\noindent The initial distribution and the transition matrices are common knowledge, whereas the realizations of valuations are private information.

The expected budget deficit is $1/4$ in $t=0$ and $(s_{00}+s_{10})(b_{01}+b_{11})/4$ in $t=1$. This is so since $\th_s^1=0$ with probability $(s_{00}+s_{10})/2$ and $\th_b^1=1$ with probability $(b_{01}+b_{11})/2$. Now consider the seller with $\th_s^0=0$. His expected payoff is $1/2$ in $t=0$ and $s_{00}(b_{01}+b_{11})/2$ in $t=1$. Likewise, the expected payoff of the seller with $\th_s^0=1$ is zero in $t=0$ and $s_{10}(b_{01}+b_{11})/2$ in $t=1$. Similarly, the expected payoff of the buyer with $\th_b^0=0$ is zero in $t=0$ and $b_{01}(s_{00}+s_{10})/2$ in $t=1$, and that with $\th_b^0=1$ is $1/2$ in $t=0$ and $b_{11}(s_{00}+s_{10})/2$ in $t=1$. Hence, budget balance can be achieved if
$$\eqalign{& \ \ \ \ \ \frac{1+(s_{00}+s_{10})(b_{01}+b_{11})}{4} \cr
\leq & \min\Bigl\{ \frac{1+s_{00}(b_{01}+b_{11})}{2}, \frac{s_{10}(b_{01}+b_{11})}{2} \Bigr\} + \min\Bigl\{\frac{b_{01}(s_{00}+s_{10})}{2}, \frac{1+b_{11}(s_{00}+s_{10})}{2} \Bigr\}.}$$

When valuations are independent across periods so that $s_{ij} = b_{ij}=1/2$ for all $i,j=0,1$, then both the LHS and the RHS are equal to 1/2. Thus, budget balance is achieved. When valuations are persistent over time so that $s_{00}=s_{11}=b_{00}=b_{11}=1$ and $s_{01}=s_{10}=b_{01}=b_{10}=0$, then the LHS is 1/2 while the RHS is 0. Thus, budget balance cannot be achieved. Another interesting case is when players are symmetric so that $P_s = P_b$ and, moreover,
$$P_s = P_b = \pmatrix{\a & 1-\a \cr
                       1-\a & \a \cr}.$$
In this case, the inequality becomes $1/2 \leq 1-\a$, i.e., $\a \leq 1/2$. Thus, valuations should not be positively serially correlated for the budget balance.

More generally, we can show that budget balance cannot be achieved when (i) $s_{00} \geq 1/2$, $s_{11} \geq 1/2$, $b_{00} \geq 1/2$, $b_{11} \geq 1/2$, and moreover, (ii) $\max\{s_{00}, s_{11}\} > 1/2$ and $\max\{b_{00}, b_{11}\} > 1/2$. First, it is easy to see that
$$\frac{s_{10}(b_{01}+b_{11})}{2} \leq \frac{1+s_{00}(b_{01}+b_{11})}{2} {\rm \ \ and \ \ } \frac{b_{01}(s_{00}+s_{10})}{2} \leq \frac{1+b_{11}(s_{00}+s_{10})}{2}$$
since $s_{10} = 1-s_{11} \leq 1/2$ and $b_{01} = 1-b_{00} \leq 1/2$. Hence, we need to have
$$\frac{1+(s_{00}+s_{10})(b_{01}+b_{11})}{4}
\leq \frac{s_{10}(b_{01}+b_{11})}{2} + \frac{b_{01}(s_{00}+s_{10})}{2}$$
for budget balance. However, observe that
$$\eqalign{&1+(s_{00}+s_{10})(b_{01}+b_{11})-2 [s_{10}(b_{01}+b_{11}) + b_{01}(s_{00}+s_{10})] \cr
           =& 1+b_{11}(s_{00}-s_{10})-b_{01}(s_{00}+3 s_{10}) > 1 + \frac{1}{2}(s_{00}-s_{10})-\frac{1}{2}(s_{00}+3 s_{10})\cr
           = & 1 -2s_{10} \geq 0,}$$
where the inequalities hold due to our assumption. Thus, budget balance cannot be achieved.

This example shows that positive serial correlation of valuations precludes budget balance. Will it be still true when the number of periods increases?

\medskip
\noindent {\sl 3.3.2. Budget balance of dynamic bilateral trading}
\smallskip

We resume back to the infinite-period setup, so that $t=0,1,2,\ldots$. Assume that both $\th_s^t$ and $\th_b^t$ take one of the values from the set $\{ v_1, \ldots, v_{K} \}$, with $v_1 < v_2 < \cdots < v_K$. As before, $a^*(\th_s^t,\th_b^t)=1$ when $\th_s^t < \th_b^t$ and $a^*(\th_s^t,\th_b^t)=0$ when $\th_s^t \geq  \th_b^t$ in an efficient decision rule. So, $z_s^t=-\th_b^t$ and $z_b^t=\th_s^t$ when $a^*=1$, and $z_s^t=z_b^t=0$ when $a^*=0$. Both players' payoffs are  $\th_b^t - \th_s^t$ when the trade occurs, and zero when the trade does not occur. Let $V$ be a $K \times K$ matrix whose $ij$-th element $v_{ij}$ is equal to $v_j -v_i$ when $j > i$ and zero otherwise.

The dynamic evolution of valuations is represented by Markov chains. Let $P_s = \bigl(s_{ij}\bigr)_{i,j = 1, \ldots, K}$ and $P_b = \bigl(b_{ij}\bigr)_{i,j = 1, \ldots, K}$ be the seller's and the buyer's transition matrix, respectively, and let $x = (x_1, \ldots, x_K)^T$ and $y = (y_1, \ldots, y_K)^T$ be the seller's and the buyer's $K \times 1$ distribution vector of initial valuation at $t=0$, respectively, where the superscript $T$ denotes the transpose.

Observe that (i) the expected budget deficit in $t=0$ is $x^T V y$, and (ii) the seller's and the buyer's distribution vector in period $t$ is $x^T P_s^t$ and $y^T P_b^t$,  respectively, where $P_s^t$ ($P_b^t$) is the $t$-th power of $P_s$ ($P_b$), and so the expected budget deficit in period $t$ is $x^T P_s^t V (P_b^t)^T y$. By defining the $K \times K$ matrix $Q^{(t)} \equiv P_s^t V (P_b^t)^T$, the expected budget deficit in the dynamic pivot mechanism is
$$\sum_{t=0}^\infty \d^t x^T Q^{(t)} y.$$

Let $e_k$ be the $K \times 1$ vector whose $k$-th element is 1 while other elements are all zero. Then, the seller's expected payoff when $\th_s^0 = v_k$ is $\sum_{t=0}^\infty \d^t e_k^T Q^{(t)} y$ and the buyer's expected payoff when $\th_b^0 = v_k$ is $\sum_{t=0}^\infty \d^t x^T Q^{(t)} e_k$. Thus, budget balance can be achieved with lump-sum fees if and only if
$$\sum_{t=0}^\infty \d^t x^T Q^{(t)} y \leq \min_{k =1, \ldots, K} \Bigl\{ \sum_{t=0}^\infty \d^t e_k^T Q^{(t)} y \Bigr\}+ \min_{k =1, \ldots, K} \Bigl\{\sum_{t=0}^\infty \d^t x^T Q^{(t)} e_k\Bigr\}.\eqno(*)$$

We discuss several special cases before presenting general results. First of all, when $P_s=P_b=I$ where $I$ is the $K \times K$ identity matrix, so that valuations are perfectly correlated across periods, we have $Q^{(t)} = V$ for all $t \geq 0$.  Thus,
$$\sum_{t=0}^\infty \d^t x^T Q^{(t)} y = \sum_{t=0}^\infty \d^t \Bigl(\sum_{i=1}^K \sum_{j=i+1}^K (v_j-v_i)x_i y_j \Bigr)=\frac{1}{1-\d}\sum_{i=1}^K \sum_{j=i+1}^K (v_j-v_i)x_i y_j.$$
We also have
$$\min_{k =1, \ldots, K} \Bigl\{ \sum_{t=0}^\infty \d^t e_k^T Q^{(t)} y \Bigr\} = \min_{k =1, \ldots, K} \sum_{t=0}^\infty \d^t v_{k\cdot} y = \sum_{t=0}^\infty \d^t v_{K\cdot} y=0$$
where $v_{k\cdot}$ is the $k$-th row of $V$. Likewise,
$$\min_{k =1, \ldots, K} \Bigl\{\sum_{t=0}^\infty \d^t x^T Q^{(t)} e_k\Bigr\}= \min_{k =1, \ldots, K} \sum_{t=0}^\infty \d^t x^T v_{\cdot k} = \sum_{t=0}^\infty \d^t x^T v_{\cdot 1}=0$$
where $v_{\cdot k}$ is the $k$-th column of $V$. Thus, budget balance cannot be achieved unless $x_i y_j$'s are all zero for $i=1, \ldots, K$ and $j=i+1, \ldots, K$.

Next, when $P_s = P_b =P$ and $P$ is the $K \times K$ matrix whose elements are all $1/K$'s, so that valuations are independent across periods, we have $P^t = P$ and $Q^{(t)} = (1/K^2) {\bf 1} V {\bf 1}$ for all $t \geq 1$ where $\bf 1$ is the $K \times K$ matrix whose elements are all 1's. Thus,
$$\sum_{t=0}^\infty \d^t x^T Q^{(t)} y =  x^TVy + \sum_{t=1}^\infty \d^t \frac{1}{K^2} \sum_{i=1}^K \sum_{j=i+1}^K (v_j - v_i).$$
We also have
$$\sum_{t=0}^\infty \d^t e_k^T Q^{(t)} y = e_k^T V y + \sum_{t=1}^\infty \d^t \frac{1}{K^2}\sum_{i=1}^K \sum_{j=i+1}^K (v_j - v_i).$$
Likewise,
$$\sum_{t=0}^\infty \d^t x^T Q^{(t)} e_k = x^T V e_k + \sum_{t=1}^\infty \d^t \frac{1}{K^2}\sum_{i=1}^K \sum_{j=i+1}^K (v_j - v_i).$$
Thus, budget balance is achieved when
$$x^TVy + \sum_{t=1}^\infty \d^t \frac{1}{K^2} \sum_{i=1}^K \sum_{j=i+1}^K (v_j - v_i) \leq 2 \sum_{t=1}^\infty \d^t \frac{1}{K^2} \sum_{i=1}^K \sum_{j=i+1}^K (v_j - v_i),$$
i.e.,
$$\sum_{i=1}^K \sum_{j=i+1}^K (v_j - v_i) x_i y_j \leq \frac{1}{K^2} \frac{\d}{1-\d} \sum_{i=1}^K \sum_{j=i+1}^K (v_j - v_i).$$
This inequality is true for large $\d$, that is, for
$$\d \geq \frac{K^2 \sum_{i=1}^K \sum_{j=i+1}^K (v_j - v_i) x_i y_j}{K^2 \sum_{i=1}^K \sum_{j=i+1}^K (v_j - v_i) x_i y_j + \sum_{i=1}^K \sum_{j=i+1}^K (v_j - v_i).}$$

Thirdly, let us continue the example in the previous subsection and study the case when (i) $K=2$ with $v_1=0$ and $v_2=1$, so players' valuations can take either zero or one, and (ii) $P_s = P_b = P$ and
$$P =\pmatrix{\a & 1-\a \cr
                       1-\a & \a \cr}.$$
We have
$$P^t = \pmatrix{\frac{1}{2}+\frac{1}{2} (2\a-1)^t & \frac{1}{2}-\frac{1}{2} (2\a-1)^t \cr
					           \frac{1}{2}-\frac{1}{2} (2\a-1)^t & \frac{1}{2}+\frac{1}{2} (2\a-1)^t \cr}
							   \equiv \pmatrix{\a^{(t)} & 1-\a^{(t)} \cr  1-\a^{(t)} & \a^{(t)} \cr}$$
and thus
$$Q^{(t)} = \pmatrix{\a^{(t)}(1-\a^{(t)}) & (\a^{(t)})^2 \cr (1-\a^{(t)})^2 & \a^{(t)}(1-\a^{(t)}) \cr}.$$
Given the initial distributions $x=(1/2,1/2)^T$ and $y=(1/2,1/2)^T$, the budget deficit is
$$\sum_{t=0}^\infty \d^t (1/2, 1/2) Q^{(t)} \pmatrix{1/2 \cr 1/2} = \frac{1}{4(1-\d)}.$$
On the other hand, the seller's expected payoff when $\th_s^0=0$ is
$$\sum_{t=0}^\infty \d^t (1,0) Q^{(t)} \pmatrix{1/2 \cr 1/2} = \sum_{t=0}^\infty \d^t \cdot \frac{\a^{(t)}}{2} = \frac{1}{4}\Bigl(\frac{1}{1-\d} + \frac{1}{1-\d(2\a-1)}\Bigr),$$
while that when $\th_s^0=1$ is
$$\sum_{t=0}^\infty \d^t (0,1) Q^{(t)} \pmatrix{1/2 \cr 1/2} = \sum_{t=0}^\infty \d^t \cdot \frac{1-\a^{(t)}}{2} = \frac{1}{4}\Bigl(\frac{1}{1-\d} - \frac{1}{1-\d(2\a-1)}\Bigr).$$
Likewise, the buyer's expected payoff when $\th_b^0=0$ is
$$\sum_{t=0}^\infty \d^t (1/2, 1/2) Q^{(t)} \pmatrix{1 \cr 0} = \frac{1}{4}\Bigl(\frac{1}{1-\d} - \frac{1}{1-\d(2\a-1)}\Bigr)$$
and that when $\th_b^0=1$ is
$$\sum_{t=0}^\infty \d^t (1/2, 1/2) Q^{(t)} \pmatrix{0 \cr 1} = \frac{1}{4}\Bigl(\frac{1}{1-\d} + \frac{1}{1-\d(2\a-1)}\Bigr).$$
Hence, budget balance is achieved if
$$\frac{1}{4(1-\d)} \leq \frac{1}{2} \Bigl(\frac{1}{1-\d} - \frac{1}{1-\d(2\a-1)}\Bigr),$$
i.e., if $\d \geq 1/(3-2\a)$. For any $\a < 1$, this inequality is satisfied for large enough $\d$. Therefore, in contrast to the two-period case, budget balance is achieved for any $\a < 1$ when periods are infinite and players are sufficiently patient.

These examples suggest that budget balance of the dynamic pivot mechanism can be achieved with lump-sum fees unless valuations are perfectly correlated across periods. Indeed, we have:

\thm5 If the Markov chains for the seller and the buyer are irreducible and aperiodic, budget balance is achieved for sufficiently large $\d$. \ok

\pf Let $s_{ij}^{(t)}$ be the $ij$-th element of $P_s^t$, and let $b_{ij}^{(t)}$ be the $ij$-th element of $P_b^t$. By the well-known facts on finite Markov chains, there is a unique stationary distribution $\mu^s$ such that (i) $s_{ij}^{(t)} \rightarrow \mu_j^s$ as $t \rightarrow \infty$ for all $i,j = 1, \ldots, K$, and (ii) $\mu_j^s > 0$ for all $j = 1, \ldots, K$. Likewise, there is unique stationary distribution $\mu^b$ such that (i) $b_{ij}^{(t)} \rightarrow \mu_j^b$ as $t \rightarrow \infty$ for all $i,j = 1, \ldots, K$, and (ii) $\mu_j^b > 0$ for all $j = 1, \ldots, K$. Thus, for any $\e > 0$, there is $t_0$ such that $|s_{ij}^{(t)} - \mu_j^s| < \e$ and $|b_{ij}^{(t)} - \mu_j^b| < \e$ for $t \geq t_0$.

Observe that
$$x^T P_s^t = \bigl(\sum_{h=1}^K x_h s_{h1}^{(t)}, \cdots, \sum_{h=1}^K x_h s_{hK}^{(t)}\bigr), {\rm \  and \ \ }  y^T P_b^t = \bigl(\sum_{h=1}^K y_h b_{h1}^{(t)}, \cdots, \sum_{h=1}^K y_h b_{hK}^{(t)}\bigr).$$
Thus, for arbitrary $x$ and $y$, we have
$$\eqalign{& x^T Q^{(t)} y = x^T P_s^t V (P_b^t)^T y = \sum_{i=1}^K \sum_{j=i+1}^K (v_j-v_i) \Bigl(\sum_{h=1}^K x_h s_{hi}^{(t)} \Bigr) \Bigl(\sum_{h=1}^K y_h b_{hj}^{(t)} \Bigr) \cr
< & \sum_{i=1}^K \sum_{j=i+1}^K (v_j-v_i)\Bigl(\sum_{h=1}^K x_h (\mu_i^s + \e)\Bigr) \Bigl(\sum_{h=1}^K y_h (\mu_j^b + \e))\Bigr) \cr
= & \sum_{i=1}^K \sum_{j=i+1}^K (v_j-v_i)(\mu_i^s + \e)(\mu_j^b + \e)}$$
for $t \geq t_0$. Hence,
$$\sum_{t=0}^\infty \d^t x^T Q^{(t)} y < \sum_{t=0}^{t_{0}-1} \d^t x^T Q^{(t)} y + \frac{\d^{t_{0}}}{1-\d}\Bigl(\sum_{i=1}^K \sum_{j=i+1}^K (v_j-v_i)(\mu_i^s + \e)(\mu_j^b + \e)\Bigr).$$
On the other hand, we have
$$e_k^T Q^{(t)} y > \sum_{i=1}^K \sum_{j=i+1}^K (v_j-v_i)(\mu_i^s - \e)\Bigl(\sum_{h=1}^K y_h(\mu_j^b - \e)\Bigr) = \sum_{i=1}^K \sum_{j=i+1}^K (v_j-v_i)(\mu_i^s - \e)(\mu_j^b - \e)$$
and
$$x^T Q^{(t)} e_k > \sum_{i=1}^K \sum_{j=i+1}^K (v_j-v_i)\Bigl(\sum_{h=1}^K x_h (\mu_i^s - \e)\Bigr)(\mu_j^b - \e)=\sum_{i=1}^K \sum_{j=i+1}^K (v_j-v_i)(\mu_i^s - \e)(\mu_j^b - \e)$$
for $t \geq t_0$. Hence,
$$\eqalign{&\min_{k} \Bigl\{ \sum_{t=0}^\infty \d^t e_k^T Q^{(t)} y \Bigr\}+ \min_{k} \Bigl\{\sum_{t=0}^\infty \d^t x^T Q^{(t)} e_k\Bigr\}
= \sum_{t=0}^\infty \d^t e_{k_{s}}^T Q^{(t)} y + \sum_{t=0}^\infty \d^t x^T Q^{(t)} e_{k_{b}} \cr
> & \sum_{t=0}^{t_{0}-1} \d^t e_{k_{s}}^T Q^{(t)} y + \sum_{t=0}^{t_{0}-1} \d^t x^T Q^{(t)} e_{k_{b}}+ \frac{2 \d^{t_{0}}}{1-\d} \Bigl(\sum_{i=1}^K \sum_{j=i+1}^K (v_j-v_i)(\mu_i^s - \e)(\mu_j^b - \e)\Bigr),}$$
where $k_s$ and $k_b$ respectively is a value that attains the minimum. Therefore,

$$\eqalign{& \min_{k =1, \ldots, K} \Bigl\{ \sum_{t=0}^\infty \d^t e_k^T Q^{(t)} y \Bigr\}+ \min_{k =1, \ldots, K} \Bigl\{\sum_{t=0}^\infty \d^t x^T Q^{(t)} e_k\Bigr\} - \sum_{t=0}^\infty \d^t x^T Q^{(t)} y \cr
 > & \sum_{t=0}^{t_{0}-1} \d^t e_{k_{s}}^T Q^{(t)} y  + \sum_{t=0}^{t_{0}-1} \d^t x^T Q^{(t)} e_{k_{b}} -  \sum_{t=0}^{t_{0}-1} \d^t x^T Q^{(t)} y \cr
+ & \frac{\d^{t_{0}}}{1-\d} \sum_{i=1}^K \sum_{j=i+1}^K (v_j-v_i)\Bigl(\mu_i^s \mu_j^b - 3\e(\mu_i^s + \mu_j^b)+\e^2 \Bigr).
}$$

Observe that, since $0 \leq x^T Q^{(t)} y \leq C < \infty$ for any $x$ and $y$, we have
$$ \sum_{t=0}^{t_{0}-1} \d^t e_{k_{s}}^T Q^{(t)} y  + \sum_{t=0}^{t_{0}-1} \d^t x^T Q^{(t)} e_{k_{b}} -  \sum_{t=0}^{t_{0}-1} \d^t x^T Q^{(t)} y \geq - \frac{1-\d^{t_{0}}}{1-\d} C.$$
Observe also that there is $\eta > 0$ such that $\mu_i^s \mu_j^b - 3\e(\mu_i^s + \mu_j^b)+\e^2 > \eta$ for sufficiently small $\e > 0$ and that $\eta$ is independent of the discount factor $\d$. Thus, as $\d \rightarrow 1$, the term $-(1-\d^{t_{0}})C/(1-\d)$ goes to $-t_{0} C$ whereas
$$\frac{\d^{t_{0}}}{1-\d} \sum_{i=1}^K \sum_{j=i+1}^K (v_j-v_i)\Bigl(\mu_i^s \mu_j^b - 3\e(\mu_i^s + \mu_j^b)+\e^2 \Bigr)$$
goes to infinity. Therefore, condition $(*)$ is satisfied and so budget balance is achieved. \endpf

Theorem 5 establishes that the dynamic pivot mechanism with lump-sum fees is ex-post efficient, periodic ex-post incentive compatible and individually rational, and ex-ante budget balancing. This was done by showing that condition $(*)$ is satisfied under appropriate assumptions on the Markov chain and the discount factor.

We next show that budget balance cannot be achieved under the diverse preference assumption of Bergemann and V\"alim\"aki (2010). The diverse preference assumption is essential in establishing that the dynamic pivot mechanism is the only efficient mechanism that satisfies ex-post incentive compatibility, ex-post participation constraint, and efficient exit condition. Thus, it is rather unfortunate that this precludes even ex-ante budget balance.

\thm6 The dynamic pivot mechanism cannot achieve budget balance under the diverse preference assumption. \ok

\pf In our environment, part (i) of the diverse preference assumption implies that the transition matrix $P_s$ is such that $s_{KK} = 1$ (while $s_{K1}=\cdots=s_{K,K-1}=0$) and the transition matrix $P_b$ is such that $b_{11}=1$ (while $b_{12}=\cdots=b_{1K}=0$).\note{Part (i) of the diverse preference assumption is as follows: For all $i$, there exists $\underline \th_i \in \Th_i$ such that for all $a$, we have $v_i(\underline \th_i, a)=0$ and $F_i(\underline \th_i; \underline \th_i, a) = 1$ where $F_i(\cdot)$ is a transition function.} It is straightforward to check that the $KK$-th element of $P_s^t$ and the $11$-th element of $P_b^t$ are also equal to 1, that is, $s_{KK}^{(t)} = 1$ (while $s_{K1}^{(t)}= \cdots = s_{K,K-1}^{(t)}=0$) and $b_{11}^{(t)} = 1$ (while $b_{12}^{(t)}= \cdots = b_{1K}^{(t)}=0$) for all $t \geq 1$. Hence,
$$\min_{k =1, \ldots, K} \Bigl\{ \sum_{t=0}^\infty \d^t e_k^T Q^{(t)} y \Bigr\}  = \sum_{t=0}^\infty \d^t e_K^T P_s^t V (P_b^t)^T y = 0$$
since $e_K^T P_s^t = e_K^T$ and $e_K^T P_s^t V = 0$, the $1 \times K$ vector whose elements are all zero.\note{Recall that $v_{ij}=0$ for $j \leq i$.} Likewise,
$$\min_{k =1, \ldots, K} \Bigl\{\sum_{t=0}^\infty \d^t x^T Q^{(t)} e_k\Bigr\} = \sum_{t=0}^\infty \d^t x^T P_s^t V (P_b^t)^T e_1 = 0.$$
On the other hand, $\sum_{t=0}^\infty \d^t x^T Q^{(t)} y > 0 $ in general. \endpf

\noindent The reason for this result is that the Markov chain is reducible under the diverse preference assumption.

We have demonstrated that (i) budget balance of the dynamic pivot mechanism can be achieved when the Markov chain is irreducible and aperiodic, and (ii) the diverse preference assumption may preclude budget balance. These results can be extended to more general environments beyond bilateral trading: See Yoon (2015).

\Section{CONCLUSION}

We have given an elementary introduction to dynamic mechanism design. We have examined both optimal dynamic mechanisms and efficient dynamic mechanisms. As for optimal dynamic mechanisms, we have found necessary and sufficient conditions for perfect Bayesian incentive compatibility and formulated the optimal dynamic mechanism problem. As for efficient dynamic mechanisms, we have established that the dynamic Groves mechanism is the only outcome efficient and periodic ex-post incentive compatible mechanism by porting the corresponding result for static mechanism design. We have also investigated budget balance of the dynamic pivot mechanism in some detail for a bilateral trading environment to understand better the role of transition kernel regarding the evolution of private information. We have demonstrated that many results and techniques of static mechanism design can be straightforwardly extended and adapted to the analysis of dynamic settings.

This paper has considered standard frameworks. We admit that some dynamic environments, such as non-Markovian dynamic environments, may require a call for novel insight and techniques. We leave it to future research work.

\ref

\paper{Baron, D., Besanko, D.}{1984}{Regulation and information in a continuing relationship}{{\it Information Economics and Policy\/} 1}{267-302}

\paper{Bergemann, D. and Pavan, A.}{2015}{Introduction to symposium on dynamic contracts and mechanism design}{\jet 159(B)}{679-701}

\paper{Bergemann, D. and Said, M.}{2010}{Dynamic auctions}{Cochran, J., Cox, L., Keskinocak, P., Kharoufeh, J., and Smith, C. (Eds.), {\it Wiley Encyclopedia of Operations Research and Management Science,\/} Wiley}{1511-1522}

\paper{Bergemann, D. and V\"alim\"aki, J.}{2010}{The dynamic pivot mechanism}{\emet 78}{771-789}

\paper{Bergemann, D. and V\"alim\"aki, J.}{2019}{Dynamic mechanism design: An introduction}{{\it Journal of Economic Literature\/} 57}{235-274}

\paper{Carbajal, J. C.}{2010}{On the uniqueness of Groves mechanisms and the payoff equivalence principle}{\geb 68}{763-772}

\paper{Cavallo, R.}{2008}{Efficiency and redistribution in dynamic mechanism design}{{\it Proceedings of the 9th ACM Conference on Electronic Commerce\/}}{220-229}

\paper{Chatterjee, K. and Samuelson, K.}{1983}{Bargaining under incomplete information}{{\it Operations Research} 31}{835-851}

\paper{d'Aspremont, C. and G\'erard-Varet, L.-A.}{1979}{Incentives and incomplete information}{\jpube 11}{25-45}

\paper{Green, J. and Laffont, J.-J.}{1977}{Characterization of satisfactory mechanisms for the revelation of preferences for public goods}{\emet 45}{427-438}

\book{Hern\'andez-Lerma, O. and Lasserre, J.}{1996}{Discrete-Time Markov Control Processes: Basic Optimality Criteria}{Springer}

\paper{Holmstr\"om, B.}{1979}{Groves' scheme on restricted domains}{\emet 47}{1137-1144}

\paper{Myerson, R.}{1981}{Optimal auction design}{{\it Mathematics of Operations Research\/} 6}{58-73}

\paper{Myerson, R. and Satterthwaite, M.}{1983}{Efficient mechanisms for bilateral trading}{\jet 29}{265-281}

\paper{Pavan, A.}{2017}{Dynamic mechanism design: Robustness and endogenous types}{Honor\'e, B, Pakes, A., Piazzesi, M., and Samuelson, L. (Eds.), {\it Advances in Economics and Econometrics: Eleventh World Congress,\/} Cambridge University Press}{1-62}

\paper{Pavan, A., Segal, I., Toikka, J.}{2014}{Dynamic mechanism design: A Myersonian approach}{\emet 82}{601-653}

\paper{Vickrey, W.}{1961}{Counterspeculation, auctions, and competitive sealed tenders}{{\it Journal of Finance\/} 16}{8-37}

\paper{Vohra, R.}{2012}{Dynamic mechanism design}{{\it Surveys in Operations Research and Management Science} 17}{60-68}

\paper{Walker, M}{1978}{A note on the characterization of mechanisms for the revelation of preferences}{\emet 46}{147-152}

\paper{Yoon, K.}{2001}{The modified Vickrey double auction}{\jet 101}{572-584}

\paper{Yoon, K.}{2008}{The participatory Vickrey-Clarke-Groves mechanism}{\jme 44}{324-336}

\paper{Yoon, K.}{2015}{On budget balance of the dynamic pivot mechanism}{\geb 94}{206-213}

\paper{Yoon, K.}{2021}{The uniqueness of dynamic Groves mechanisms on restricted domains}{{\it Korean Economic Review\/}}{forthcoming}

\bye